\documentclass[12pt]{article}
\usepackage[top=25truemm,bottom=27truemm,left=22truemm,right=22truemm]{geometry}

\usepackage{indentfirst}
\usepackage[dvipdfmx]{graphicx}
\usepackage{float}
\usepackage{url}
\usepackage{siunitx}
\usepackage{array}
\usepackage{amsmath}
\usepackage{amssymb}
\usepackage{fancybox}
\usepackage{ascmac}
\usepackage{bm}
\usepackage{mathrsfs}
\usepackage{cancel}
\usepackage{physics}
\usepackage{color}
\usepackage{authblk}
\usepackage{appendix}
\usepackage{comment}
\usepackage{mhchem}
\usepackage{caption,latexsym,amsfonts,mathtools,here}
\usepackage{subcaption}
\usepackage{hyperref}
\usepackage{multirow}

\numberwithin{equation}{section}

\newcommand{\no}{\nonumber}

\newcommand{\hhh}{\hspace{6mm}}

\newcommand{\pa}{\partial}

\newcommand{\eqns}{\hspace{-5pt}&=&\hspace{-5pt}}
\newcommand{\defs}{\hspace{-5pt}&\equiv&\hspace{-5pt}}
\newcommand{\beqn}{\begin{eqnarray}}
\newcommand{\eeqn}{\end{eqnarray}}

\begin{document}
\begin{titlepage}
\begin{flushright}
{YITP-25-104, KOBE-COSMO-25-09}
\end{flushright}

\vspace{50pt}

\begin{center}

{\large{\textbf{Parity violation in photon quasinormal modes of black holes }}}

\vspace{25pt}

{Sugumi Kanno$^{*\natural}$, Jiro Soda$^{\flat}$, and Akira Taniguchi$^*$}
\end{center}

\vspace{20pt}

\shortstack[l]
{\hspace{1.8cm}\it {\small $^*$Department of Physics, Kyushu University, Fukuoka 819-0395, Japan} \\[5pt]
\hspace{1.8cm}\it {\small $^\natural$Center for Gravitational Physics and Quantum Information,} \\[2pt]
\hspace{1.93cm}\it {\small Yukawa Institute for Theoretical Physics, Kyoto University,} \\[2pt]
\hspace{1.8cm}\it {\small $^\flat$Department of Physics, Kobe University, Kobe 657-8501, Japan}}
\vspace{28pt}

\begin{abstract}

Given that black holes are ubiquitous in the universe and axion-like scalar fields are potential candidates for dark energy and/or dark matter, it is natural to consider cosmological black holes endowed with axion. We investigate the photon quasinormal modes of a Schwarzschild black hole with axion where the electromagnetic field is coupled to the axion field via a Chern-Simons interaction. We derive the master equations for the electromagnetic field as a set of coupled equations for parity-even and parity-odd modes and numerically compute quasinormal modes by using Leaver's continued fraction method. We find parity violation in the polarization of photons within the quasinormal mode spectrum. This parity violation in electromagnetic signals could serve as a new probe to explore the nature of the dark sector.
\end{abstract}
\end{titlepage}

\tableofcontents

\section{Introductioin}

Parity violation is a key concept in particle physics. 
Remarkably, it may also appear in the gravity sector.
For instance, primordial gravitational waves may be circularly polarized~\cite{Satoh:2007gn,Satoh:2008ck,Takahashi:2009wc}.
The four-point correlation function of galaxies may exhibit parity violation~\cite{Philcox:2022hkh}. Since such effects could offer deep
physical insight into the nature of gravity, it is worthwhile to broaden the search for phenomena that exhibit parity violation. In this paper, we focus on quasinormal modes of black holes.

An important issue in cosmology is to clarify the nature of the dark sector, namely, dark matter and dark energy. Historically, Weakly Interacting Massive Particles (WIMPs) have been among the most promising candidates for dark matter. However, since no evidence for supersymmetry has been discovered at the LHC, it is important to remain open-minded in our research for the true nature of the dark sector. In fact, axions, originally introduced to resolve the strong CP problem~\cite{Peccei:1977hh, Weinberg:1977ma, Wilczek:1977pj, Kim:1979if, Shifman:1979if, Dine:1981rt, Zhitnitsky:1980tq} have become canidates for dark matter~\cite{Preskill:1982cy, Abbott:1982af, Dine:1982ah, Hui:2016ltb}.  Moreover, axion-like particles (ALPs), pseudo-scalar fields predicted by string theory~\cite{Svrcek:2006yi}, are being intensively studied as candidates for both dark matter and dark energy~\cite{Arvanitaki:2009fg}.
This is because ALPs can have a wide range of masses. Depending on their mass, axions can behave either as  scalar dark matter or as dark energy. For example, axions with masses in the range from $10^{-22}\sim 10^{3}$~eV oscillate coherently~\cite{Turner:1983he,Hu:2000ke,Hui:2016ltb} and behave like dark matter. On the other hand, axions with the masses around $10^{-33}$~eV, comparable to the current Hubble scale, can act as dark energy~\cite{Amendola:2005ad}. 

Black holes are ubiquitous in the universe. In fact, a supermassive black hole is found at the center of a galaxy, and LIGO has discovered binary black holes with masses of several tens of solar masses. Interestingly, primordial black holes (PBHs), which are formed from density fluctuations in the early universe, have been proposed as a candidate for dark matter~\cite{Khlopov:2008qy,Carr:2021bzv}. The mass range in which PBHs could account for a significant fraction of dark matter is $10^{16}\sim 10^{22}$~g \cite{Carr:2021bzv}. Hawking radiation from such ultralight PBHs might be observable, as the Hawking temperature is inversely proportional to the black hole mass~\cite{Hawking:1974rv}. Therefore, observations of X-rays and gamma-rays produced by Hawking radiation could lead to the detection of PBHs or give constraints on their abundance. 

As mentioned above, ALPs are predicted in superstring theory and Black holes can form through both astrophysical and cosmological process. 
Thus, it is natural to consider ALPs and PBHs in a unified framework. 
Several studies have investigated scalar fields around black holes. In \cite{Jacobson:1999vr},   the effect of a massless scalar field on the evolution of PBHs was examined within the framework of scalar-tensor gravity. It was shown that linearly time-dependent scalar field solutions in Schwarzschild spacetime can endow black holes with scalar field. Furthermore, in \cite{Hui:2019aqm}, the authors studied the profile of massive scalar field dark matter in Schwarzschild spacetime. 

Given that axion fields can endow black holes with,  it is natural to investigate how such axion affects the surrounding electromagnetic field.  In this paper, we focus on the photon quasinormal modes of  black hole with axion. Axions and electromagnetic fields are coupled through the Chern-Simons (CS) term, which is known to induce various phenomena, such as circular polarization, amplification due to instabilities, and modification of speed of light via changes in the dispersion relation~\cite{Yoshida:2017ehj,Arza:2018dcy}. We investigate parity violation in photon quasinormal modes of black holes endowed with axion. 

The paper is organized as follows. In Section 2, we derive the master equations for electromagnetic field coupled to the axion field in Schwarschild spacetime. In Section 3, we present the master equations for photons in a black hole spacetime with the time dependent axion background. We also discuss the conditions for the existence of quasinormal modes. In section 4, we derive the recursion relations and evaluate quasinormal modes. As a consequence, we find the parity violation in the spectrum of quasinormal modes. 
The final section is devoted to conclusion. 
We gather matrix elements in Appendix A.
We work in the natural and geometrized units: $\hbar=c=G=1$.

\section{Master equations for photon quasinormal modes}
In cosmology, black holes and scalar fields are ubiquitous. In particular,
an axion like scalar field could be dark matter or dark energy depending on its mass. The axion field interacts with electromagnetic fields through the Chern-Simons term. Hence, in this section, we derive master equations for electromagnetic fields in black holes endowed with axion.

We consider  the Schwarzschild black hole
    \begin{eqnarray}
	ds^2 \eqns -\left(1-\frac{2M}{r}\right)dt^2+\left(1-\frac{2M}{r}\right)^{-1}dr^2+r^2 d\theta^2+r^2\sin^2\theta d\varphi^2\no\\[6pt]
	\defs -f(r)dt^2+f(r)^{-1}dr^2+r^2 d\theta^2+r^2\sin^2\theta d\varphi^2\ .
	\end{eqnarray}
and a cosmological axion field $\phi$ as a background.  On this background, we 
consider propagation of electromagnetic waves. 
The action of the electromagnetic field $A_\mu$ coupled to the axion field $\phi$ is given by
    \begin{eqnarray}
    \label{eq:action}
	S\eqns \int d^4x \sqrt{-g}\left[-\frac{1}{4}F_{\mu\nu}F^{\mu\nu}-\frac{1}{4}\alpha \phi F_{\mu\nu}\tilde{F}^{\mu\nu}\right]\ ,
	\end{eqnarray}
where $F_{\mu\nu}$ represents the field strength defined as $F_{\mu\nu}=\pa_\mu A_\nu-\pa_\nu A_\mu$. Here, $\tilde{F}^{\mu\nu}=\frac{1}{2}\varepsilon^{\mu\nu\rho\sigma}F_{\rho\sigma}$ ($\sqrt{-g}\  \epsilon^{0123} =-1$) is a dual of  $F_{\mu\nu}$ and $\alpha$ represents the coupling constant between axion field and electromagnetic field. Since axion is a real pseudoscalar, the only way to retain the parity symmetry is to use the above action~(\ref{eq:action}). The variation of the action (\ref{eq:action}) with respect to the gauge field $A_\mu$ gives the Maxwell equation:
    \begin{eqnarray}
    \label{eq:maxwell}
	\nabla_\mu F^{\mu\nu}+\alpha\varepsilon^{\mu\nu\rho\sigma}(\nabla_\mu \phi)\nabla_\rho A_\sigma = 0\ .
	\end{eqnarray}
In order to solve Maxwell equation (\ref{eq:maxwell}) in spherically symmetric spacetime, we write equation Eq.~(\ref{eq:maxwell}) in polar coordinates. 
Note that the relation
\begin{eqnarray}
  r^2 \sin \theta \epsilon^{0rJK} = - f^{-1/2} r^2 \sin\theta\epsilon^{rJK} 
  =  - r^2 \sin\theta\epsilon^{JK} 
\end{eqnarray}
leads to
\begin{eqnarray}
   \epsilon^{0rJK}   = - \epsilon^{JK} \ .
\end{eqnarray}
Here, the capital indices $J$ and $K$ take the angular part of the spherical coordinates $\theta$ and $\varphi$, and $\epsilon^{JK}$ is a Levi-civita tensor on the sphere.
Assuming that the axion only depends on the time $t$ and the radial coordinate $r$, we obtain
    \begin{eqnarray}
	&&\nabla_\mu F^{\mu t}+\alpha\phi'\varepsilon^{IJ}\nabla_I A_J = 0\ ,\label{eq:1}\\[2pt]
	&&\nabla_\mu F^{\mu r}-\alpha\dot{\phi}\varepsilon^{IJ}\nabla_I A_J = 0\ ,\label{eq:2}\\[2pt]
	&&\nabla_\mu F^{\mu I}-\alpha\dot{\phi}\varepsilon^{IJ}\left(\nabla_J A_r-\nabla_r A_J\right)+\alpha\phi'\varepsilon^{I J}\left(\nabla_J A_t-\nabla_t A_J\right) =  0\ ,\label{eq:3}
	\end{eqnarray}
where a dot denotes the derivative with respect to $t$, and a prime denotes the derivative with respect to $r$. To represent the electromagnetic field in polar coordinates, we expand the gauge field with spherical harmonics $Y_{lm}(\theta, \varphi)$ as follows.
    \begin{eqnarray}
	A_i \eqns \sum_{l m}a_{i, l m}(t,r)Y_{lm}(\theta, \varphi)\ ,\label{eq:Arho}\\[4pt]
	A_I \eqns \sum_{l m}\left[a^{(+)}_{l m}(t,r)\nabla_I Y_{lm}(\theta, \varphi)+a^{(-)}_{l m}(t,r)\varepsilon_{I}{}^J\nabla_J Y_{lm}(\theta, \varphi)\right]\ ,\label{eq:AI}
	\end{eqnarray}
where the indice $i$ takes $t$ and $r$. $\varepsilon_{IJ}$ is the antisymmetric Levi-Civita tensor on 2-dimensional sphere defined as $\varepsilon_{\theta\varphi}=r^2\sin\theta$, $\varepsilon_{\varphi\theta}=-r^2\sin\theta$. In the following, the indices $l,m$ of the $a_i, a_I$ is omitted for simplicity.
Note that the vector $A_I$ is decomposed into the sum of a gradient and the rotation parts. Then $a^{(-)}$ is a gauge invariant but $a_t$, $a_r$, and $a^{(+)}$ are not. Substituting Eq.~(\ref{eq:Arho}) and Eq.~(\ref{eq:AI}) into Eq.~(\ref{eq:1}) $\sim$ (\ref{eq:3}), we obtain
    \begin{eqnarray}
	&&\hspace{-24pt}\frac{1}{r^2}\pa_r \left[r^2\left(\pa_t a_r-\pa_r a_t\right)\right]-\frac{1}{f}\frac{l(l+1)}{r^2}\left(\pa_t a^{(+)}-a_t\right)+\alpha\phi'\frac{l(l+1)}{r^2}a^{(-)}= 0\,,\label{eq1}\\[6pt]
	&&\hspace{-24pt}-\pa_t \left(\pa_t a_r-\pa_r a_t\right)+f\frac{l(l+1)}{r^2}\left(\pa_r a^{(+)}-a_r\right)-\alpha\dot{\phi}\frac{l(l+1)}{r^2}a^{(-)}= 0\,,\label{eq2}\\[6pt]
	&&\hspace{-24pt}-\pa_t\left(\pa_t a^{(+)}-a_t\right)+f\pa_r\left[f\left(\pa_r a^{(+)}-a_r\right)\right]-f\alpha\dot{\phi}\pa_r a^{(-)}+f\alpha\phi'\pa_t a^{(-)} = 0\,,\label{eq3}\\[6pt]
	&&\hspace{-24pt}-\pa^2_t a^{(-)}+f\pa_r\left(f\pa_r a^{(-)}\right)-f\frac{l(l+1)}{r^2}a^{(-)}+f\alpha\dot{\phi}\left(\pa_r a^{(+)}-a_r\right)-f\alpha\phi'\left(\pa_t a^{(+)}-a_t\right) = 0\,.\no\\ \label{eq4}
	\end{eqnarray}
Note that the combination $\pa_t a_r-\pa_r a_t=F_{tr}$ is gauge invariant. Then we introduce gauge invariant variables $\Psi^{(+)}$ and $\Psi^{(-)}$ as follows 
    \begin{eqnarray}
	\pa_t a_r-\pa_r a_t \eqns \frac{l(l+1)}{r^2}\Psi^{(+)}\ ,\label{eqBdef}\\[6pt]
	a^{(-)}\eqns \Psi^{(-)}\ ,\label{eqAdef}
	\end{eqnarray}
where $\Psi^{(+)}$ is the even-parity mode and $\Psi^{(-)}$ is the odd parity mode. 
Substituting Eq.~(\ref{eqBdef}) and Eq.~(\ref{eqAdef}) into Eq.~(\ref{eq1}) and Eq.~(\ref{eq2}), we obtain the following equations
    \begin{eqnarray}
	f\pa_r \Psi^{(+)} - \left(\pa_t a^{(+)}-a_t\right)+f\alpha\phi'\Psi^{(-)} \eqns 0\,, \label{eq1-2}\\[6pt]
	-\frac{1}{f}\pa_t \Psi^{(+)} + \left(\pa_r a^{(+)}-a_r\right)-\frac{1}{f}\alpha\dot{\phi}\Psi^{(-)} \eqns 0\,. \label{eq2-2}
	\end{eqnarray}
We then see that Eq.~(\ref{eq3}) is automatically satisfied by using Eqs.~(\ref{eq1-2}) and (\ref{eq2-2}). In order to express~Eqs.~(\ref{eq1-2}) and (\ref{eq2-2}) using only gauge invariant variables, we take a combination $\pa_t$(\ref{eq2-2})$+$$\pa_r$(\ref{eq1-2}) to have 
    \begin{eqnarray}
	-\frac{\pa^2 \Psi^{(+)}}{\pa t^2}+f\frac{\pa}{\pa r}\left(f\frac{\pa\Psi^{(+)}}{\pa r}\right)-V(r)\Psi^{(+)}-\alpha\frac{\pa}{\pa t}\left(\dot{\phi}\Psi^{(-)}\right)+f\alpha\frac{\pa}{\pa r}\left(f\phi'\Psi^{(-)}\right)= 0\,.\label{eq:comp1}
	\end{eqnarray}
Here, we used Eq.~(\ref{eqBdef}) and $V(r)$ is effective potential defined as
    \begin{eqnarray}
	V(r) \equiv f\frac{l(l+1)}{r^2}\ .\label{effpotential}
	\end{eqnarray}
Similarly, using Eqs.~(\ref{eq1-2}) and~(\ref{eq2-2}), Eq.~(\ref{eq4}) becomes gauge invariant form expressed by
    \begin{eqnarray}
	&&\hspace{-24pt}-\frac{\pa^2 \Psi^{(-)}}{\pa t^2}+f\frac{\pa}{\pa r}\left(f\frac{\pa \Psi^{(-)}}{\pa r}\right)-V(r)\Psi^{(-)}+\left[\left(\alpha\dot{\phi}\right)^2-\left(f\alpha \phi'\right)^2\right] \Psi^{(-)}\no\\[4pt]
    &&\hspace{7cm}+\alpha\dot{\phi}\frac{\pa \Psi^{(+)}}{\pa t}-f^2\alpha\phi'\frac{\pa \Psi^{(+)}}{\pa r} = 0\,.\label{eq:comp2}
	\end{eqnarray}
Remarkably, two parity modes $\Psi^{(+)}$ and $\Psi^{(-)}$ are mixed through the Chern-Simons coupling. 

For later convenience, we introduce the tortoise coorinate $r^{*}$, which satisfies $dr/dr^{*}=f$, Eqs.~(\ref{eq:comp1}) and (\ref{eq:comp2}) become
    \begin{eqnarray}
	&&\hspace{-24pt}-\frac{\pa^2 \Psi^{(+)}}{\pa t^2}+\frac{\pa^2 \Psi^{(+)}}{\pa {r^*}^2}-V(r)\Psi^{(+)}-\alpha\frac{\pa}{\pa t}\left(\dot{\phi}\Psi^{(-)}\right)+\alpha\frac{\pa}{\pa r^*}\left(\frac{\pa \phi}{\pa r^*}\Psi^{(-)}\right)= 0\ ,\label{eq:ev}\\[6pt]
	&&\hspace{-24pt}-\frac{\pa^2 \Psi^{(-)}}{\pa t^2}+\frac{\pa^2 \Psi^{(-)}}{\pa {r^*}^2}-V(r)\Psi^{(-)}+\left[\left(\alpha\dot{\phi}\right)^2-\left(\alpha \frac{\pa \phi}{\pa r^*}\right)^2\right]\Psi^{(-)} \no\\[4pt]
    &&\hspace{6cm}+ \alpha\dot{\phi}\frac{\pa \Psi^{(+)}}{\pa t}-\alpha\frac{\pa \phi}{\pa r^*}\frac{\pa \Psi^{(+)}}{\pa r^*}= 0\ .\label{eq:od}
	\end{eqnarray}
\section{Photons in the black hole background with axion}
\subsection{Cosmological axion field}

Since the time scales between the black hole and the cosmological background are separated, it is legitimate to think the black hole
lies in a local asymptotically flat space.
However, we need to take into account the boundary condition for the axion field determined by the cosmological evolution. 
The large separation of scales allows us to consider the following cosmological boundary condition
    \begin{eqnarray}
    \label{eq:phi01}
	\phi(t)\big|_{r\rightarrow \infty}=\phi_{\rm c} t,\hhh \phi_{\rm c}=\rm const.
	\end{eqnarray}
Note that as long as $\phi_{\rm c}\ll 1/M$, the backreaction to Schwarzschild spacetime is negligible.
The Klein-Gorden equation is
    \begin{eqnarray}
    \label{eq:masslessKlein}
	\left(-g^{tt}\pa^2_t+\frac{1}{\sqrt{-g}}\pa_r\sqrt{-g}g^{rr}\pa_r\right)\phi(t, r)=0\ .
	\end{eqnarray}
The solution under the boundary condition (\ref{eq:phi01}) is found in~\cite{Jacobson:1999vr} of the form
    \begin{eqnarray}
	\phi(t,r)=\phi_{\rm c}\left[t+2M\ln\left(1-\frac{2M}{r}\right)\right],\hhh \phi_{\rm c}=\rm const.\label{eq:phi02}
	\end{eqnarray}
The point here is that the solution does not diverge at the horizon. This can be found by introducing Eddington-Finkelstein coordinate $v=t+r^*$. 
Using Eddington-Finkelstein time which is regular at the horizon, 
    \begin{eqnarray}
	v=t+r+2M\ln\left(\frac{r}{2M}-1\right)\ ,
	\end{eqnarray}
The solution~(\ref{eq:phi02}) can be written as
    \begin{eqnarray}
    \tilde{\phi}(v,r)={\phi}(t,r)
    \eqns\phi_{\rm c}\left[v-r-2M\ln\left(\frac{r}{2M}\right)\right]\ .
	\end{eqnarray}
Since $v$ is regular at the horizon, we see that the scalar field $\phi(t,r)$ is also regular at the horizon. Thus, plugging the above solution into  (\ref{eq:comp1}) and (\ref{eq:comp2}), the master equations for photons in the black hole background become
    \begin{eqnarray}
	&&\hspace{-24pt}-\frac{\pa^2 \Psi^{(+)}}{\pa t^2}+f\frac{\pa}{\pa r}\left(f\frac{\pa\Psi^{(+)}}{\pa r}\right)-V\Psi^{(+)}-\alpha\phi_{\rm c}\frac{\pa\Psi^{(-)}}{\pa t}+f\alpha\phi_{\rm c}\frac{\pa}{\pa r}\left(\frac{(2M)^2}{r^2}\Psi^{(-)}\right)= 0\ ,\label{eq:mastereq1}\\[8pt]
    &&\hspace{-24pt}-\frac{\pa^2 \Psi^{(-)}}{\pa t^2}+f\frac{\pa}{\pa r}\left(f\frac{\pa \Psi^{(-)}}{\pa r}\right)-V\Psi^{(-)}+\left(\alpha\phi_{\rm c}\right)^2\left[1-\left(\frac{2M}{r}\right)^4\right] \Psi^{(-)}\no\\[4pt]
	&&\hspace{6cm}+\alpha\phi_{\rm c}\frac{\pa \Psi^{(+)}}{\pa t}-f\alpha\phi_{\rm c}\frac{(2M)^2}{r^2}\frac{\pa \Psi^{(+)}}{\pa r} = 0\ .\label{eq:mastereq2}
	\end{eqnarray}
By using $dr/dr^{*}=f$, Eqs.~(\ref{eq:mastereq1}) and (\ref{eq:mastereq2}) in the tortoise coodinate $r^{*}$ are expressed as
    \begin{eqnarray}
	&&\hspace{-24pt}-\frac{\pa^2 \Psi^{(+)}}{\pa t^2}+\frac{\pa}{\pa r^*}\left(\frac{\pa\Psi^{(+)}}{\pa r^*}\right)-V\Psi^{(+)}-\alpha\phi_{\rm c}\frac{\pa\Psi^{(-)}}{\pa t}+\alpha\phi_{\rm c}\frac{\pa}{\pa r^*}\left(\frac{(2M)^2}{r^2}\Psi^{(-)}\right)= 0\ ,\label{eq:mastereq3}\\[8pt]
    &&\hspace{-24pt}-\frac{\pa^2 \Psi^{(-)}}{\pa t^2}+\frac{\pa}{\pa r^*}\left(\frac{\pa \Psi^{(-)}}{\pa r^*}\right)-V\Psi^{(-)}+\left(\alpha\phi_{\rm c}\right)^2\left[1-\left(\frac{2M}{r}\right)^4\right] \Psi^{(-)}\no\\[4pt]
	&&\hspace{6cm}+\alpha\phi_{\rm c}\frac{\pa \Psi^{(+)}}{\pa t}-\alpha\phi_{\rm c}\frac{(2M)^2}{r^2}\frac{\pa \Psi^{(+)}}{\pa r^*} = 0\ .\label{eq:mastereq4}
	\end{eqnarray}

\subsection{Effective potential}
Quasinormal modes refer to solutions that satisfy the outgoing boundary condition at infinity and the ingoing boundary condition at the horizon. Therefore, when considering a quasinormal mode, there must be a local maximum point in the effective potential. In Eq.~(\ref{eq:mastereq2}), the effective potential of $\Psi^{(-)}$ is expressed as
    \beqn
    V_{\rm eff}^{\left(\Psi^{(-)}\right)} = f\frac{l(l+1)}{r^2}-\left(\alpha\phi_{\rm c}\right)^2\left[1-\left(\frac{2M}{r}\right)^4\right]\ ,
    \eeqn
where we used Eq.~({\ref{effpotential}) and the interaction terms with respect to $\Psi^{(+)}$ are ignored for simplicity. The condition for this potential to have a local maximum
    \beqn
    0=\frac{d}{dr} V_{\rm eff}^{\left(\Psi^{(-)}\right)}= \frac{2}{r^2}\frac{l(l+1)}{r^2}-\frac{2}{r}f\frac{l(l+1)}{r^2}-\frac{64\left(\alpha\phi_{\rm c}\right)^2}{r^5}\ ,
    \eeqn
and the stationary point $r_0$ is is found to be
    \beqn
    r_0=\frac{6Ml(l+1) + \sqrt{[6Ml(l+1)]^2 - 512l(l+1)M^4 \left(\alpha\phi_{\rm c}\right)^2}}{4l(l+1)}\ .
    \eeqn
In order for the stationary point to be outside the horizon $r_0>2M$, we need the following condition
    \begin{eqnarray}
        \alpha\phi_{\rm c} < \frac{\sqrt{l(l+1)}}{4M}\ .\label{eq:alphacon}
    \end{eqnarray}
This is because there is no stationary point unless the above is satisfied, which means that quasinormal modes do not exist. In the next section, we evaluate the quasinormal modes in the region~(\ref{eq:alphacon}).

\section{Parity violation in photon quasinormal modes}
\subsection{Asymptotic solutions}
In this subsection, we derive asymptotic solutions for $\Psi^{(\pm)}$ near the horizon and infinity.

We consider a stationary state solution $\Psi\sim e^{-i\omega t}$ in the tortoise coordinate. Eqs.~(\ref{eq:mastereq3}) and (\ref{eq:mastereq4}) near the horizon~($r^*\to-\infty$) are expressed as
    \begin{eqnarray}
    \omega^2 \Psi^{(+)} + \frac{\partial^2 \Psi^{(+)}}{\partial r^{*2} }+ i \omega \alpha \phi_{\rm c}\Psi^{(-)}+\alpha\phi_{\rm c}\frac{\partial}{\partial r^{*}} \Psi^{(-)} \eqns 0\ , \label{eq:4.1.1}\\[4pt]
    \omega^2 \Psi^{(-)} + \frac{\partial^2  \Psi^{(-)}}{\partial r^{*2}} -i\omega\alpha\phi_{\rm c}\Psi^{(+)}- \alpha \phi_{\rm c} \frac{\partial}{\partial r^*} \Psi^{(+)} \eqns 0\ .\label{eq:4.1.2}
    \end{eqnarray}
The ingoing boundary condition near the horizon is expressed as
    \begin{eqnarray}
    \Psi(r^*) \propto e^{-i k r^*}, \quad k > 0\ .
    \end{eqnarray}
After substituting this into Eqs.~(\ref{eq:4.1.1}) and (\ref{eq:4.1.2}), we consider the conditions required for the existence of a non-trivial solution,
    \begin{eqnarray}
    \det\left(\begin{array}{cc}
    \omega^2 - k^2 & i \omega \alpha \phi_{\rm c} -i k \alpha \phi_{\rm c} \\
    -i \omega \alpha \phi_{\rm c} + ik \alpha \phi_{\rm c} & \omega^2 - k^2
    \end{array}\right) = 0\ .
    \end{eqnarray}
Then the wavenumber near the horizon turns out to be 
    \begin{eqnarray} 
    k^{\rm H}_{\pm}\eqns \frac{1}{2} \left( \pm\alpha \phi_{\rm c} + \sqrt{(2\omega\mp \alpha\phi_{\rm c})^2} \right)=\omega \label{kpm-horizon}\ ,
    \end{eqnarray}
where we chose the positive branch of the square root, that is, $\sqrt{z^2}=+z$. Then the asymptotic solutions near the horizon are represented by the following linear combination of eigenvectors
    \begin{align}
    \begin{pmatrix} \Psi^{(+)} \\[4pt] \Psi^{(-)}\end{pmatrix} \sim\begin{pmatrix} 1 \\ i \end{pmatrix} e^{-i k^{\rm H}_+ r^*} + \begin{pmatrix} 1 \\ -i \end{pmatrix} e^{-i k^{\rm H}_- r^*}\ . \label{eq:asy01}
    \end{align}

On the other hand, in the limit at infinity~($r^* \to \infty$),  Eqs.~(\ref{eq:mastereq3}) and (\ref{eq:mastereq4}) become
    \begin{eqnarray}
    \omega^2 \Psi^{(+)} + \frac{\partial^2}{\partial r^{*2}} \Psi^{(+)} +i \omega \alpha \phi_{\rm c} \Psi^{(-)} \eqns 0\ , \\
    \omega^2 \Psi^{(-)} + \frac{\partial^2}{\partial r^{*2}} \Psi^{(-)} + (\alpha \phi_{\rm c})^2 \Psi^{(-)} - i\omega \alpha \phi_{\rm c} \Psi^{(+)} \eqns 0\ .
    \end{eqnarray}
The outgoing boundary condition at infinity is expressed as
    \begin{eqnarray}
    \Psi(r^*) &\sim& e^{i k r^*}, \hhh k>0\ .
    \end{eqnarray}
Substituting this outgoing solution into Eqs.~(\ref{eq:4.1.1}) and (\ref{eq:4.1.2}), we have  the conditions for a non-trivial solution
    \begin{eqnarray}
    \det\begin{pmatrix}
    \omega^2 - k^2 & i \omega \alpha \phi_{\rm c} \\
    -i \omega \alpha \phi_{\rm c} & \omega^2 - k^2 + (\alpha \phi_{\rm c})^2
    \end{pmatrix} &=& 0\,,
    \end{eqnarray}
and this leads to 
    \begin{eqnarray}
    k^\infty_{\pm} &=& \frac{1}{2} \left( \mp\alpha \phi_{\rm c} + \sqrt{\left(\alpha \phi_{\rm c}\right)^2+ 4\omega^2} \right)\,.
    \end{eqnarray}
The asymptotic solutions at infinity are represented by the following linear combination,
    \begin{eqnarray}
    \begin{pmatrix} \Psi^{(+)} \\ \Psi^{(-)}\end{pmatrix} \sim \begin{pmatrix} 1 \\ i\frac{k^\infty_+}{\omega} \end{pmatrix} e^{i k^\infty_+ r^*}+ \begin{pmatrix} 1 \\ -i\frac{k^\infty_-}{\omega}  \end{pmatrix} e^{i k^\infty_- r^*} \ . \label{eq:asy02}
    \end{eqnarray}
\subsection{Recurrence relations}
Next, we calculate the quasinormal modes using Leaver's continued fraction method \cite{Leaver:1985ax}. First, we perform the separation of variables on Eqs.~(\ref{eq:mastereq1}) and (\ref{eq:mastereq2}) such as
    \beqn
	\Psi^{(\pm)}(t,r)=\frac{1}{r}\psi^{(\pm)}(r, \omega)e^{-i\omega t}\ ,\label{eq:furidef}
	\eeqn
and we define the following variables
	\beqn
	\kappa=-2iM\omega,\hhh  x=\frac{r}{2M}\ .
	\eeqn
Then, Eqs.~(\ref{eq:mastereq1}) and (\ref{eq:mastereq2}) are written as
	\beqn
	&&\hspace{-24pt}\left[x(x-1)\frac{\pa}{\pa x^2}+\frac{\pa}{\pa x}-\left(\frac{\kappa^2 x^3}{x-1}+l(l+1)\right)\right]\psi^{(+)}+i\omega\alpha\phi_{\rm c} \psi^{(-)}+\frac{\alpha\phi_{\rm c}}{2M}\frac{x-1}{x}\frac{\pa}{\pa x}\left(\frac{1}{x^2}\psi^{(-)}\right)=0\ ,\label{eq:mas01}\no\\\\
	&&\hspace{-24pt}\left[x(x-1)\frac{\pa}{\pa x^2}+\frac{\pa}{\pa x}-\left(\frac{\kappa^2 x^3}{x-1}+l(l+1)\right)\right]\psi^{(-)}+\left(\alpha\phi_{\rm c}\right)^2\left[1-\frac{1}{x^4}\right]\psi^{(-)}\no\\[8pt]
	&&\hspace{7cm}-i\omega\alpha\phi_{\rm c}\psi^{(+)}-\frac{\alpha\phi_{\rm c}}{2M}\frac{x-1}{x^3}\frac{\pa}{\pa x}\psi^{(+)}=0\ .\label{eq:mas02}
	\eeqn
Next, we define $\rho_{\rm H/I}$ and $\sigma_{\rm H/I}$ near the horizon and at infinity respectively such as
	\beqn
	\rho_{\rm H}=-2Mik^{\rm H}_+,\hhh  \sigma_{\rm H}=-2Mik^{\rm H}_-,\hhh  \rho_{\rm I}=-2Mik^\infty_+,\hhh  \sigma_{\rm I}=-2Mik^\infty_-\ .
	\eeqn
The relation between the tortoise coordinate $r^*$ and $x$ is written by
    \begin{eqnarray}
    r^*=r+2M\ln\left(\frac{r}{2M}-1\right)=2M\left[x+\ln\left(x-1\right)\right] \ .
    \end{eqnarray}
Then, the asymptotic solutions near the horizon~(\ref{eq:asy01}) and at infinity~(\ref{eq:asy02}) are written in terms of $\rho_{H/I}, \sigma_{H/I}$ and $x$, respectively as follows
    \begin{eqnarray}
    \begin{pmatrix} \psi^{(+)} \\ \psi^{(-)}\end{pmatrix} \hspace{-5pt}&\sim &\hspace{-5pt} \begin{pmatrix} 1 \\ i \end{pmatrix} (x-1)^{\rho_{\rm H}} + \begin{pmatrix} 1 \\ -i \end{pmatrix} (x-1)^{\sigma_{\rm H}}\ , \label{eq:hor}\\[4pt]
    \begin{pmatrix} \psi^{(+)} \\ \psi^{(-)}\end{pmatrix} \hspace{-5pt}&\sim &\hspace{-5pt} \begin{pmatrix} 1 \\ -\frac{\rho_{\rm I}}{2M\omega} \end{pmatrix} e^{-\rho_{\rm I} x}x^{-\rho_{\rm I}} + \begin{pmatrix} 1 \\ \frac{\sigma_{\rm I}}{2M\omega}  \end{pmatrix} e^{-\sigma_{\rm I} x}x^{-\sigma_{\rm I}}\ .\label{eq:inf}
\end{eqnarray}
Thus, the solution which satisfies the boundary condition near the horizon and at infinity can be written as
	\begin{eqnarray}
	\psi^{(+)}\eqns (x-1)^{\rho_{\rm H}} e^{-\rho_{\rm I}(x-1)}x^{-(\rho_{\rm I}+\rho_{\rm H})}\sum_{n=0}^\infty a_n \left(\frac{x-1}{x}\right)^n\no\\[4pt]
	&&+(x-1)^{\sigma_{\rm H}} e^{-\sigma_{\rm I}(x-1)}x^{-(\sigma_{\rm I}+\sigma_{\rm H})}\sum_{n=0}^\infty b_n \left(\frac{x-1}{x}\right)^n\ , \label{eq:psipl}\\[8pt]
	\psi^{(-)}\eqns -\frac{\rho_{\rm I}}{2M\omega}(x-1)^{\rho_{\rm H}} e^{-\rho_{\rm I}(x-1)}x^{-(\rho_{\rm I}+\rho_{\rm H})}	\sum_{n=0}^\infty a_n \left(\frac{x-1}{x}\right)^n\no\\[4pt]
	&&+\frac{\sigma_{\rm I}}{2M\omega}(x-1)^{\sigma_{\rm H}} e^{-\sigma_{\rm I}(x-1)}x^{-(\sigma_{\rm I}+\sigma_{\rm H})}\sum_{n=0}^\infty b_n \left(\frac{x-1}{x}\right)^n\ . \label{eq:psimi}
	\end{eqnarray}
In the limit $x\to 1$, Eq.~(\ref{eq:psipl}) and (\ref{eq:psimi}) become
	\begin{eqnarray}
    \lim_{x\to 1}\psi^{(+)} \eqns (x-1)^{-2Mi \omega}(a_0 + b_0)\ ,\\[4pt]
	\lim_{x\to 1}\psi^{(-)} \eqns (x-1)^{-2Mi \omega}\left(-\frac{\rho_{\rm I}}{2M\omega}a_0+\frac{\sigma_{\rm I}}{2M\omega}b_0\right)\ .
	\end{eqnarray}
Since Eq.~(\ref{eq:hor}) and (\ref{eq:inf}) show that $\psi^{(+)}=(x-1)^{-2Mi\omega}$ and $\psi^{(-)}=0$ near the horizon, we choose $a_0$ and $b_0$ to satisfy 
    \begin{eqnarray}
    \left\{ \,
    \begin{aligned}
    & a_0 + b_0 = 1 \\[4pt]
    & -\frac{\rho_{\rm I}}{2M\omega}a_0+\frac{\sigma_{\rm I}}{2M\omega}b_0 = 0 
    \end{aligned}
    \right.
    \hhh\to\hhh
    \left\{ \,
    \begin{aligned}
    & a_0 =\frac{\sigma_{\rm I}}{\rho_{\rm I}+\sigma_{\rm I}} \\[4pt]
    & b_0 =\frac{\rho_{\rm I}}{\rho_{\rm I}+\sigma_{\rm I}} 
    \end{aligned}
    \right.\ ,
    \end{eqnarray}

Substituting Eqs.~(\ref{eq:psipl}) and (\ref{eq:psimi}) into Eqs.~(\ref{eq:mas01}) and (\ref{eq:mas02}) and defining $\bm{c}_n$ as 
    \beqn
    \bm{c}_{n}=
	\begin{pmatrix}
   	a_n \\
   	b_n
	\end{pmatrix} \ ,
    \eeqn
we obtain the following recurrence relation.  
    \beqn
    &&\mathcal{A}_{n}^{(8)}\bm{c}_{1}+\mathcal{B}_{n}^{(8)}\bm{c}_{0}=0\\[2pt]
    &&\mathcal{A}_{n}^{(8)}\bm{c}_{2}+\mathcal{B}_{n}^{(8)}\bm{c}_{1}+\mathcal{C}_{n}^{(8)}\bm{c}_{0}=0\\[2pt]
    &&\mathcal{A}_{n}^{(8)}\bm{c}_{3}+\mathcal{B}_{n}^{(8)}\bm{c}_{2}+\mathcal{C}_{n}^{(8)}\bm{c}_{1}+\mathcal{D}_{n}^{(8)}\bm{c}_{0}=0\\[2pt]
    &&\mathcal{A}_{n}^{(8)}\bm{c}_{4}+\mathcal{B}_{n}^{(8)}\bm{c}_{3}+\mathcal{C}_{n}^{(8)}\bm{c}_{2}+\mathcal{D}_{n}^{(8)}\bm{c}_{1}+\mathcal{E}_{n}^{(8)}\bm{c}_{0}=0\\[2pt]
    &&\mathcal{A}_{n}^{(8)}\bm{c}_{5}+\mathcal{B}_{n}^{(8)}\bm{c}_{4}+\mathcal{C}_{n}^{(8)}\bm{c}_{3}+\mathcal{D}_{n}^{(8)}\bm{c}_{2}+\mathcal{E}_{n}^{(8)}\bm{c}_{1}+\mathcal{F}_{n}^{(8)}\bm{c}_{0}=0\\[2pt]
    &&\mathcal{A}_{n}^{(8)}\bm{c}_{6}+\mathcal{B}_{n}^{(8)}\bm{c}_{5}+\mathcal{C}_{n}^{(8)}\bm{c}_{4}+\mathcal{D}_{n}^{(8)}\bm{c}_{3}+\mathcal{E}_{n}^{(8)}\bm{c}_{2}+\mathcal{F}_{n}^{(8)}\bm{c}_{1}+\mathcal{G}_{n}^{(8)}\bm{c}_{0}=0\\[2pt]
    &&\hspace{-12pt} {\rm for}~n\geq 6\no\\
    &&\mathcal{A}_{n}^{(8)}\bm{c}_{n+1}+\mathcal{B}_{n}^{(8)}\bm{c}_{n}+\mathcal{C}_{n}^{(8)}\bm{c}_{n-1}+\mathcal{D}_{n}^{(8)}\bm{c}_{n-2}+\mathcal{E}_{n}^{(8)}\bm{c}_{n-3}+\mathcal{F}_{n}^{(8)}\bm{c}_{n-4}+\mathcal{G}_{n}^{(8)}\bm{c}_{n-5}+\mathcal{H}_{n}^{(8)}\bm{c}_{n-6}=0\no\label{eq:asyeq08}\\
	\eeqn
Here, the superscript (8) indicates that the recurrence relation is between eight terms. Also, matrixs $\mathcal{A}_{n}^{(8)}\sim \mathcal{H}_{n}^{(8)}$ are defined as
	\beqn
	&&\hspace{-18pt}\mathcal{A}_{n}^{(8)}=
	\begin{pmatrix}
   	\alpha^{(\rho)}_n & \alpha^{(\sigma)}_n \\
   	A^{(\rho)}_n & A^{(\sigma)}_n
	\end{pmatrix},\ 
	\mathcal{B}_{n}^{(8)}=
	\begin{pmatrix}
   	\beta^{(\rho)}_n & \beta^{(\sigma)}_n \\
   	B^{(\rho)}_n & B^{(\sigma)}_n
	\end{pmatrix},\ 
	\mathcal{C}_{n}^{(8)}=
	\begin{pmatrix}
   	\gamma^{(\rho)}_n & \gamma^{(\sigma)}_n \\
   	C^{(\rho)}_n & C^{(\sigma)}_n
	\end{pmatrix},\ 
	\mathcal{D}_{n}^{(8)}=
	\begin{pmatrix}
   	\delta^{(\rho)}_n & \delta^{(\sigma)}_n \\
   	D^{(\rho)}_n & D^{(\sigma)}_n
	\end{pmatrix}\ ,\no\\[8pt]
	&&\hspace{-18pt}\mathcal{E}_{n}^{(8)}=
	\begin{pmatrix}
   	\epsilon^{(\rho)}_n & \epsilon^{(\sigma)}_n \\
   	E^{(\rho)}_n & E^{(\sigma)}_n
	\end{pmatrix},\ 
	\mathcal{F}_{n}^{(8)}=
	\begin{pmatrix}
   	\zeta^{(\rho)}_n & \zeta^{(\sigma)}_n \\
   	F^{(\rho)}_n & F^{(\sigma)}_n
	\end{pmatrix},\ 
	\mathcal{G}_{n}^{(8)}=
	\begin{pmatrix}
   	\eta^{(\rho)}_n & \eta^{(\sigma)}_n \\
   	G^{(\rho)}_n & G^{(\sigma)}_n
	\end{pmatrix},\ 
	\mathcal{H}_{n}^{(8)}=
	\begin{pmatrix}
   	\theta^{(\rho)}_n & \theta^{(\sigma)}_n \\
   	H^{(\rho)}_n & H^{(\sigma)}_n
	\end{pmatrix}\ .\no\\
	\eeqn
and matrix elements are listed in Appendix A. The point is that 
we can reduce the 8-term recurrence relation~(\ref{eq:asyeq08}) to a 3-term recurrence relation by following the method in~\cite{Pani:2013pma}. 

As an illustration, we present the procedure to reduce the 8-term recurrence relation~(\ref{eq:asyeq08}) to a 7-term recurrence relation: 
    \beqn
    \mathcal{A}_{n}^{(7)}\bm{c}_{n+1}+\mathcal{B}_{n}^{(7)}\bm{c}_{n}+\mathcal{C}_{n}^{(7)}\bm{c}_{n-1}+\mathcal{D}_{n}^{(8)}\bm{c}_{n-2}+\mathcal{E}_{n}^{(7)}\bm{c}_{n-3}+\mathcal{F}_{n}^{(7)}\bm{c}_{n-4}+\mathcal{G}_{n}^{(7)}\bm{c}_{n-5}=0\ .\label{eq:ac7ter}
    \eeqn
Note that the superscript $(k)$ represents the coefficients of the $k$-term recurrence relation. To transform Eq.~(\ref{eq:asyeq08}) into Eq.~(\ref{eq:ac7ter}), it is sufficient to perform the following substitutions.
	\beqn
	0\leq n\leq 5\no\\
	\mathcal{A}_{n}^{(7)}\eqns\mathcal{A}_{n}^{(8)},
    \mathcal{B}_{n}^{(7)}=\mathcal{B}_{n}^{(8)},\  \mathcal{C}_{n}^{(7)}=\mathcal{C}_{n}^{(8)},\  \mathcal{D}_{n}^{(7)}=\mathcal{D}_{n}^{(8)},\  \mathcal{E}_{n}^{(7)}=\mathcal{E}_{n}^{(8)},\ 
    \mathcal{F}_{n}^{(7)}=\mathcal{F}_{n}^{(8)},\ 
    \mathcal{G}_{n}^{(7)}=\mathcal{G}_{n}^{(8)}\no\\[8pt]
	n\geq 6\no\\
	\mathcal{A}_{n}^{(7)}\eqns\mathcal{A}_{n}^{(8)}\ ,\no\\[6pt]
	\mathcal{B}_{n}^{(7)}\eqns\mathcal{B}_{n}^{(8)}-\mathcal{H}_{n}^{(8)}\left[\mathcal{G}_{n-1}^{(7)}\right]^{-1}\mathcal{A}_{n-1}^{(7)}\ , \hhh
	\mathcal{C}_{n}^{(7)}=\mathcal{C}_{n}^{(8)}-\mathcal{H}_{n}^{(8)}\left[\mathcal{G}_{n-1}^{(7)}\right]^{-1}\mathcal{B}_{n-1}^{(7)}\ ,\no\\[4pt]
	\mathcal{D}_{n}^{(7)}\eqns\mathcal{D}_{n}^{(8)}-\mathcal{H}_{n}^{(8)}\left[\mathcal{G}_{n-1}^{(7)}\right]^{-1}\mathcal{C}_{n-1}^{(7)}\ ,\hhh
	\mathcal{E}_{n}^{(7)}=\mathcal{E}_{n}^{(8)}-\mathcal{H}_{n}^{(8)}\left[\mathcal{G}_{n-1}^{(7)}\right]^{-1}\mathcal{D}_{n-1}^{(7)}\ ,\no\\[4pt]
    \mathcal{F}_{n}^{(7)}\eqns\mathcal{F}_{n}^{(8)}-\mathcal{H}_{n}^{(8)}\left[\mathcal{G}_{n-1}^{(7)}\right]^{-1}\mathcal{E}_{n-1}^{(7)}\ ,\hhh
    \mathcal{G}_{n}^{(7)}=\mathcal{G}_{n}^{(8)}-\mathcal{H}_{n}^{(8)}\left[\mathcal{G}_{n-1}^{(7)}\right]^{-1}\mathcal{F}_{n-1}^{(7)}\ .
	\eeqn
This is because the left hand side of 8-term recurrence relation takes the form
    \beqn
	&&\mathcal{A}_{n}^{(8)} \bm{c}_{n+1}+\mathcal{B}_{n}^{(8)}\bm{c}_{n}+\mathcal{C}_{n}^{(8)}\bm{c}_{n-1}+\mathcal{D}_{n}^{(8)} \bm{c}_{n-2}+\mathcal{E}_{n}^{(8)}\bm{c}_{n-3}+\mathcal{F}_{n}^{(8)}\bm{c}_{n-4}+\mathcal{G}_{n}^{(8)}\bm{c}_{n-5}+\mathcal{H}_{n}^{(8)}\bm{c}_{n-6}\no\\[8pt]
	&&=\mathcal{A}_{n}^{(8)} \bm{c}_{n+1}+\mathcal{B}_{n}^{(8)}\bm{c}_{n}+\mathcal{C}_{n}^{(8)}\bm{c}_{n-1}+\mathcal{D}_{n}^{(8)} \bm{c}_{n-2}+\mathcal{E}_{n}^{(8)}\bm{c}_{n-3}+\mathcal{F}_{n}^{(8)}\bm{c}_{n-4}+\mathcal{G}_{n}^{(8)}\bm{c}_{n-5}\no\\[4pt]
	&&\hspace{12pt} -\mathcal{H}_{n}^{(8)}\left[\mathcal{G}_{n-1}^{(7)}\right]^{-1}\left(\mathcal{A}_{n-1}^{(7)}\bm{c}_n+\mathcal{B}_{n-1}^{(7)}\bm{c}_{n-1}+\mathcal{C}_{n-1}^{(7)}\bm{c}_{n-2}+\mathcal{D}_{n-1}^{(7)}\bm{c}_{n-3}+\mathcal{E}_{n-1}^{(7)}\bm{c}_{n-4}+\mathcal{F}_{n-1}^{(7)}\bm{c}_{n-5}\right)\no\\[8pt]
	&&=\mathcal{A}_{n}^{(7)} \bm{c}_{n+1}+\left(\mathcal{B}_{n}^{(8)}-\mathcal{H}_{n}^{(8)}\left[\mathcal{G}_{n-1}^{(7)}\right]^{-1}\mathcal{A}_{n-1}^{(7)}\right)\bm{c}_n+\left(\mathcal{C}_{n}^{(8)}-\mathcal{H}_{n}^{(8)}\left[\mathcal{G}_{n-1}^{(7)}\right]^{-1}\mathcal{B}_{n-1}^{(7)}\right)\bm{c}_{n-1}\no\\[2pt]
	&&\hspace{2.1cm} +\left(\mathcal{D}_{n}^{(8)}-\mathcal{H}_{n}^{(8)}\left[\mathcal{G}_{n-1}^{(7)}\right]^{-1}\mathcal{C}_{n-1}^{(7)}\right)\bm{c}_{n-2}+\left(\mathcal{E}_{n}^{(8)}-\mathcal{H}_{n}^{(8)}\left[\mathcal{G}_{n-1}^{(7)}\right]^{-1}\mathcal{D}_{n-1}^{(7)}\right)\bm{c}_{n-3}\no\\[2pt]
    &&\hspace{2.1cm} +\left(\mathcal{F}_{n}^{(8)}-\mathcal{H}_{n}^{(8)}\left[\mathcal{G}_{n-1}^{(7)}\right]^{-1}\mathcal{E}_{n-1}^{(7)}\right)\bm{c}_{n-2}+\left(\mathcal{G}_{n}^{(8)}-\mathcal{H}_{n}^{(8)}\left[\mathcal{G}_{n-1}^{(7)}\right]^{-1}\mathcal{F}_{n-1}^{(7)}\right)\bm{c}_{n-3}\no\\[8pt]
	&&=\mathcal{A}_{n}^{(7)} \bm{c}_{n+1}+\mathcal{B}_{n}^{(7)}\bm{c}_{n}+\mathcal{C}_{n}^{(7)}\bm{c}_{n-1}+\mathcal{D}_{n}^{(7)} \bm{c}_{n-2}+\mathcal{E}_{n}^{(7)}\bm{c}_{n-3}+\mathcal{F}_{n}^{(7)}\bm{c}_{n-4}+\mathcal{G}_{n}^{(7)}\bm{c}_{n-5}\ .
	\eeqn
Here, in the first transformation we used the following relationship derived from Eq.~(\ref{eq:ac7ter}),
    \beqn
	\left[\mathcal{G}_{n-1}^{(7)}\right]^{-1}\left(\mathcal{A}_{n-1}^{(7)}\bm{c}_{n}+\mathcal{B}_{n-1}^{(7)}\bm{c}_{n-1}+\mathcal{C}_{n-1}^{(7)}\bm{c}_{n-2}+\mathcal{D}_{n-1}^{(7)}\bm{c}_{n-3}+\mathcal{E}_{n-1}^{(7)}\bm{c}_{n-4}+\mathcal{F}_{n-1}^{(7)}\bm{c}_{n-5}\right)\eqns -\bm{c}_{n-6}\ .\no\\
	\eeqn
Thus, we have shown that the recurrence relation has been reduced to a 7-term form. 

By repeatedly performing the above procedure, we can reduce the number of terms in the recurrence relation one by one. Eventually, we obtain the 3-term recurrence relation:
	\beqn
	&&\mathcal{A}_{0}^{(3)}\bm{c}_{1}+\mathcal{B}_{0}^{(3)}\bm{c}_{0}=0\ ,\label{eq31}\\[4pt]
	&&\mathcal{A}_{n}^{(3)}\bm{c}_{n+1}+\mathcal{B}_{n}^{(3)}\bm{c}_{n}+\mathcal{C}_{n}^{(3)}\bm{c}_{n-1}=0\ .\label{eq:32}
	\eeqn
    
Finally, we transform this 3-term recurrence relation into an algebraic equation. We define the the ladder matrix $\bm{R}_n^{+}$ as
	\beqn
	\bm{c}_{n+1}=\bm{R}_{n}^+\bm{c}_n\ .\label{eq:ladder}
	\eeqn
Then, from Eqs.(\ref{eq:32}) and (\ref{eq:ladder}) we get the following expression:
	\beqn
	\bm{R}_{n}^+\eqns-\left[\mathcal{B}_{n+1}^{(3)}+\mathcal{A}_{n+1}^{(3)}\bm{R}_{n+1}^+\right]^{-1}\mathcal{C}_{n+1}^{(3)}\ . \label{eq:ladder02}
	\eeqn
Hence, we have 
	\beqn
	0\eqns \mathcal{A}_{0}^{(3)}\bm{R}_0^{+}+\mathcal{B}_{0}^{(3)}\no\\[4pt]
	\eqns\mathcal{B}_{0}^{(3)}-\mathcal{A}_{0}^{(3)}\left[\mathcal{B}_{1}^{(3)}+\mathcal{A}_{1}^{(3)}\bm{R}_{1}^+\right]^{-1}\mathcal{C}_{1}^{(3)}\no\\[4pt]
	\eqns \mathcal{B}_{0}^{(3)}-\mathcal{A}_{0}^{(3)}\left[\mathcal{B}_{1}^{(3)}-\mathcal{A}_{1}^{(3)}\left[\mathcal{B}_{2}^{(3)}+\mathcal{A}_{2}^{(3)}\bm{R}_{2}^+\right]^{-1}\mathcal{C}_{2}^{(3)}\right]^{-1}\mathcal{C}_{1}^{(3)}\no\\[4pt]
	\eqns \cdots \ .
	\eeqn
Thus, the condition for this equation to have a non-trivial solution is 
	\beqn
	\det(\mathcal{A}_{0}^{(3)}\bm{R}_0^{+}+\mathcal{B}_{0}^{(3)})=0\ .
	\eeqn
In order to determine the matrix $\bm{R}_0^{+}$, set $\bm{R}_N^{+}$ for a sufficiently large integer $N$, and then, using Eq.~(\ref{eq:ladder02}), iteratively compute $\bm{R}_{N-1}^{+}$, $\bm{R}_{N-2}^{+}$, and so on.
The procedure can continue until we reach $\bm{R}^+_0$.

\subsection{Quasinormal modes}
In this section, we present fundamental modes of quasinormal modes. In Table~\ref{tab:QNMs}, the fundamental modes are shown both for the case without coupling between the photon and the scalar field ($M\alpha\phi_{\rm c}=0$) and for the case with the coupling ($M\alpha\phi_{\rm c}=0.03$). Here, quasinormal modes are separated into their real and imaginary parts as
\beqn
    \omega=\omega_R + i\omega_I \ .
\eeqn
As seen from Eq.~(\ref{eq:furidef}), a negative imaginary part indicates that the mode amplitude decays in time. In Table~\ref{tab:QNMs}, we can check that the result for the uncoupled case ($M\alpha\phi_{\rm c}=0$) consistently reproduces the known quasinormal modes of spin 1 perturbations in Schwarzschild black hole~\cite{Mamani:2022akq}. Importantly, as we see from the results for the coupled case ($M\alpha\phi_{\rm c}=0.03$), the coupling with the axion field causes the mode to split into two branches. This is nothing but the parity violation. To see the parity violation more clearly, we plotted the $\alpha\phi_{\rm c}$ dependence of the branch of the fundamental modes in Fig.~\ref{fig:split}. Note that the range of $\alpha\phi_{\rm c}$ is constrained as $\alpha\phi_{\rm c}<\sqrt{l(l+1)}/4M$ as in Eq.~(\ref{eq:alphacon}). From Fig.~\ref{fig:split}, we see that the photon quasinormal modes coupled to a scalar field via the Chern-Simons interaction in a Schwarzschild black hole background exhibit parity violation. From Fig.~\ref{fig:split}, we see that the parity violation occurs irrespective of storength of the coupling $\alpha\phi_c$.
Curiously, there is a flip in the real part of eigenmodes (Left panels). 

To see this parity violation, conservatively the frequency of a photon should exceed $10$MHz. 
This frequency corresponds to a black hole with the mass $10^{31}$g. Therefore, if black holes with the mass less than $10^{31}$g have axion, we would be able to observe parity violation from the polarization.
The polarization must carry the information of the dark sector. Hence, parity violation in electromagnetic signals serve as a new probe to explore the nature of
the dark sector.

\begin{table}[t]
\centering
\begin{tabular}{|c|cc|cc|}
\hline
 & \multicolumn{2}{c|}{$M\alpha\phi_{\rm c} = 0$} & \multicolumn{2}{c|}{$M\alpha\phi_{\rm c} = 0.03$} \\
\cline{2-5}
 & $M\omega_R$ & $M\omega_I$ & $M\omega_R$ & $M\omega_I$ \\
\hline
\multirow{2}{*}{$l=1$} & \multirow{2}{*}{0.248263} & \multirow{2}{*}{-0.0924877} & 0.243212 & -0.0867778 \\
 & & & 0.254193 & -0.0978863 \\\hline
\multirow{2}{*}{$l=2$} & \multirow{2}{*}{0.457596} & \multirow{2}{*}{-0.0950044} & 0.457078 & -0.0930082 \\
 & & & 0.458191 & -0.0970523 \\\hline
\multirow{2}{*}{$l=3$} & \multirow{2}{*}{0.656899} & \multirow{2}{*}{-0.0956162} & 0.657182 & -0.0943654 \\
 & & & 0.656623 & -0.0969067 \\\hline
\multirow{2}{*}{$l=4$} & \multirow{2}{*}{0.853095} & \multirow{2}{*}{-0.0958599} & 0.853663 & -0.0949144 \\
 & & & 0.852521 & -0.0968368 \\\hline
\multirow{2}{*}{$l=5$} & \multirow{2}{*}{1.047913} & \multirow{2}{*}{-0.0959817} & 1.048616 & -0.0951969 \\
 & & & 1.047199 & -0.0967931 \\
\hline
\end{tabular}
\caption{The fundamental modes with and without coupling are listed.
The splitting eigenvalues indicate parity violation.}
\label{tab:QNMs}
\end{table}


\begin{figure}[htbp]
  \centering

  \begin{subfigure}[b]{\linewidth}
    \raisebox{8\height}{\makebox[0.1\linewidth][r]{\text{$l=1$}}}%
    \hspace{24pt}%
    \begin{minipage}[b]{0.65\linewidth}
      \includegraphics[width=\linewidth]{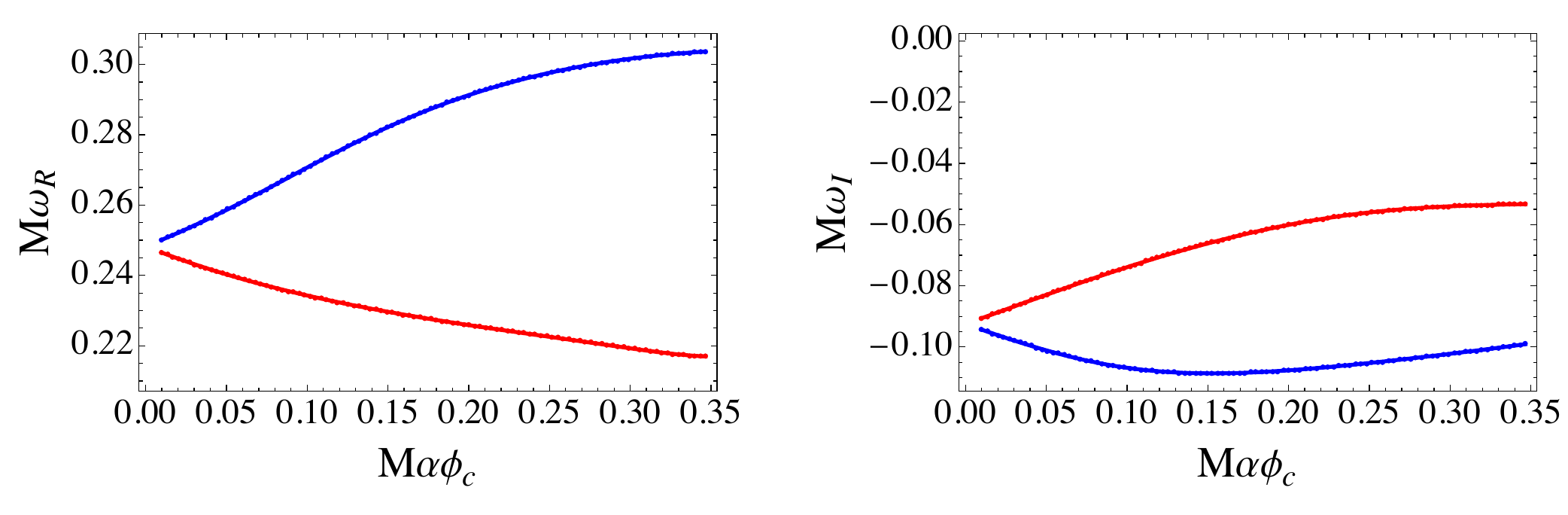}
    \end{minipage}
  \end{subfigure}

  \begin{subfigure}[b]{\linewidth}
    \raisebox{8\height}{\makebox[0.1\linewidth][r]{\text{$l=2$}}}%
    \hspace{24pt}%
    \begin{minipage}[b]{0.65\linewidth}
      \includegraphics[width=\linewidth]{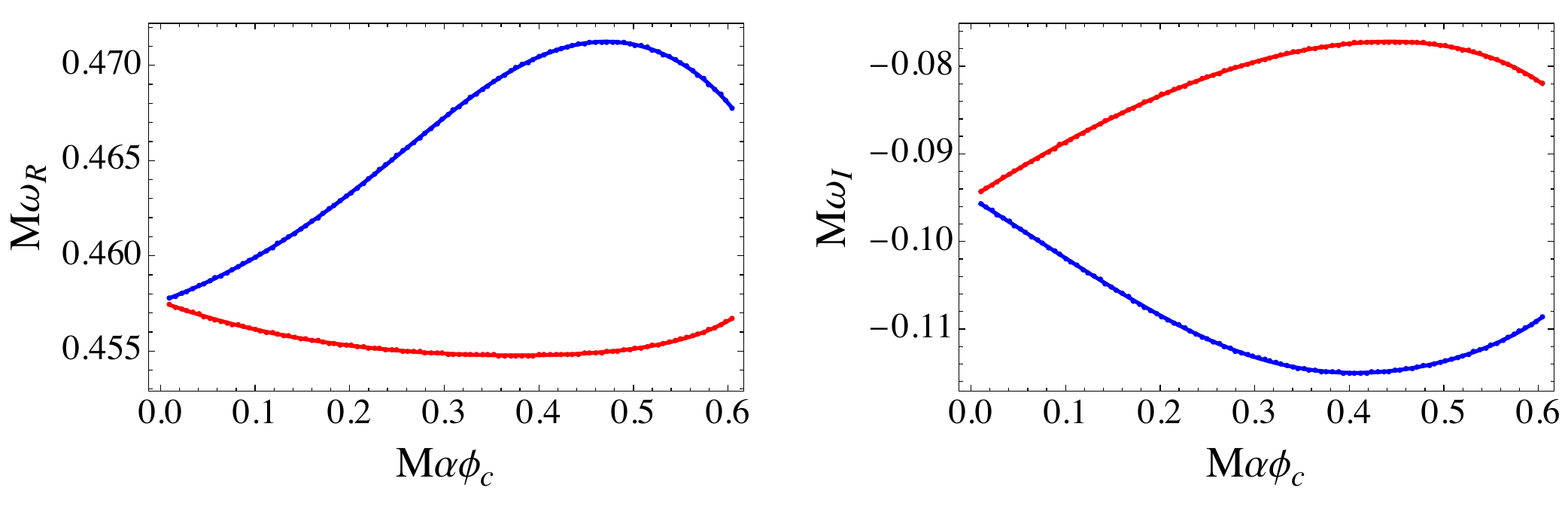}
    \end{minipage}
  \end{subfigure}

  \begin{subfigure}[b]{\linewidth}
    \raisebox{8\height}{\makebox[0.1\linewidth][r]{\text{$l=3$}}}%
    \hspace{24pt}%
    \begin{minipage}[b]{0.65\linewidth}
      \includegraphics[width=\linewidth]{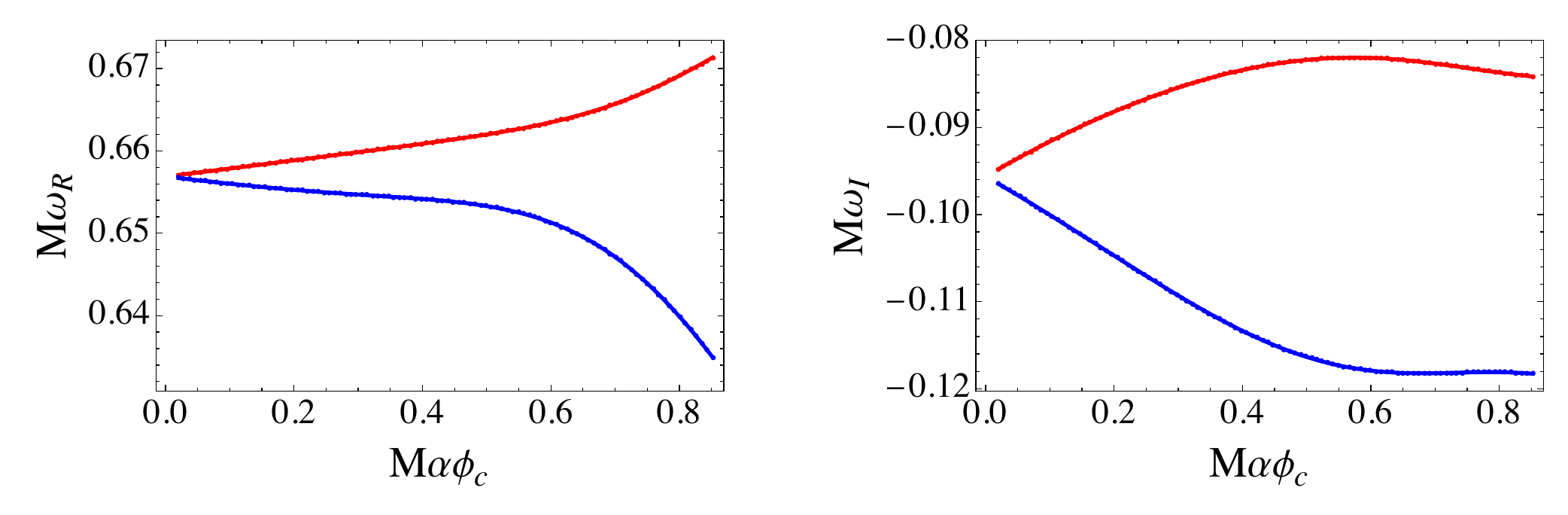}
    \end{minipage}
  \end{subfigure}

  \begin{subfigure}[b]{\linewidth}
    \raisebox{8\height}{\makebox[0.1\linewidth][r]{\text{$l=4$}}}%
    \hspace{24pt}%
    \begin{minipage}[b]{0.65\linewidth}
      \includegraphics[width=\linewidth]{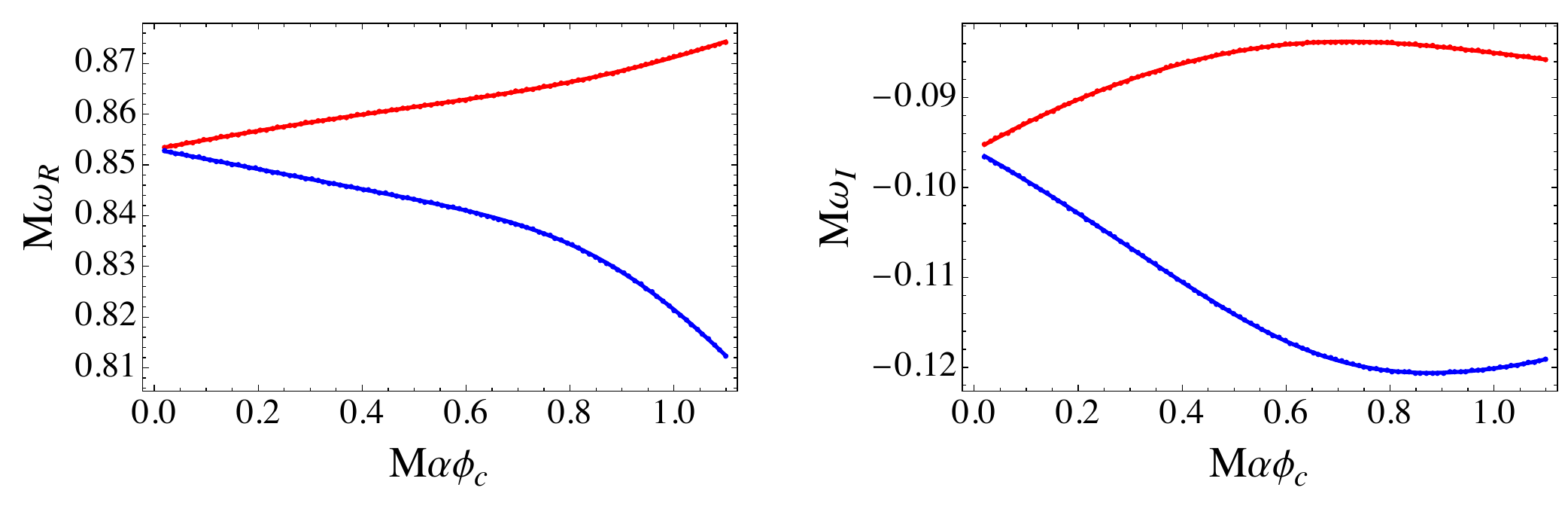}
    \end{minipage}
  \end{subfigure}

  \begin{subfigure}[b]{\linewidth}
    \raisebox{8\height}{\makebox[0.1\linewidth][r]{\text{$l=5$}}}%
    \hspace{24pt}%
    \begin{minipage}[b]{0.65\linewidth}
      \includegraphics[width=\linewidth]{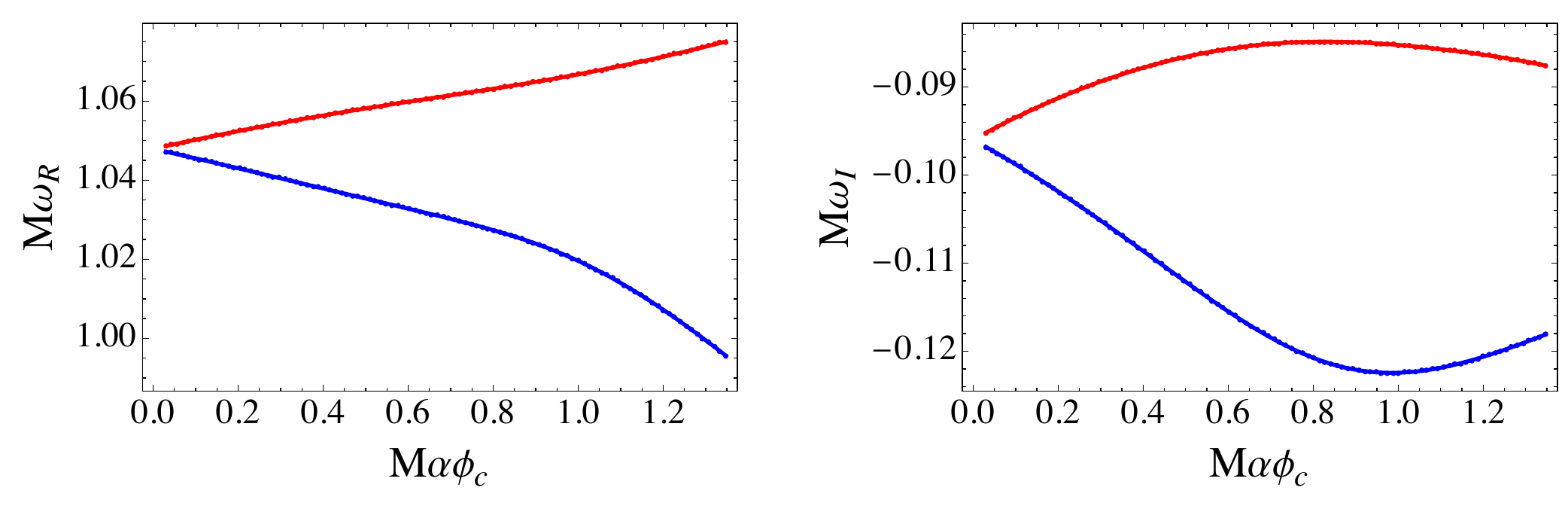}
    \end{minipage}
  \end{subfigure}

  \caption{The fundamental modes are plotted as a function of the coupling constant. The range of $\alpha\phi_{\rm c}$ is restricted as $\alpha\phi_{\rm c}<\sqrt{l(l+1)}/4M$ by Eq.~(\ref{eq:alphacon}). The modes with larger imaginary parts $ \omega_I$ are plotted with red lines, while those with smaller $ \omega_I$ are plotted with blue lines. In the absence of coupling ($\alpha\phi_{\rm c}=0$), these modes are degenerated, but as the coupling increases, we see that the quasinormal modes exhibit parity violation.}
  \label{fig:split}
\end{figure}

\section{Conclusion}
In this paper, we calculated the photon quasinormal modes of the Schwarzschild black holes with axion. We first derived the master equations for the electromagnetic field coupled to an axion field that remains regular at the black hole horizon, via a Chern-Simons (CS) interaction. We found that the resulting master equations show mixing between parity-even and parity-odd gauge-invariant variables.
Furthermore, we argued that quasinormal modes exist only when the CS coupling constant $\alpha$ is smaller than a threshold $\sqrt{l(l+1)}/4M$, which is determined by the condition for the existence of extreme values of the effective potential for the parity-odd mode. We then numerically calculated the quasinormal modes using Leaver's continued fraction method. As a consequence, we found parity violation in the quasinormal mode spectrum.
These results suggest that observations of parity-violating phenomena in the electromagnetic field could be used as potential probes of the dark sector. 

In this work, we focused on the electromagnetic sector
\begin{eqnarray}
	S\eqns \int d^4x \sqrt{-g}\left[-\frac{1}{4}F_{\mu\nu}F^{\mu\nu}
    -\frac{1}{4}\alpha \phi \ \epsilon^{\mu\nu\lambda\rho}F_{\mu\nu} F_{\lambda\rho}\right]\ .
\end{eqnarray}
However, it is known that extending the model to the gravity sector induces
interesting phenomena~\cite{Yoshida:2017cjl,Kanno:2023kdi}.
More precisely, we can caculate quasinormal modes in the following model
\begin{eqnarray}
	S\eqns \int d^4x \sqrt{-g}\left[\frac{1}{4}R-\frac{1}{4}\alpha \phi \ 
    \epsilon^{\mu\nu\lambda\rho} R_{\mu\nu\alpha\beta} R_{\lambda\rho}{}^{\alpha\beta}\right]\ .
\end{eqnarray}
Extending our analysis to gravitational waves and searching for parity violation in their signals will be reported separately~\cite{Taniguchi}.

\section*{Acknowledgments}
S.\ K. was supported by the Japan Society for the Promotion of Science (JSPS) KAKENHI Grant Numbers JP22H01220, 24K21548 and MEXT KAKENHI Grant-in-Aid for Transformative Research Areas A “Extreme Universe” No. 24H00967.
J.\ S. was in part supported by JSPS KAKENHI Grant Numbers JP23K22491, JP24K21548, JP25H02186. A.\ T. was supported by JSPS KAKENHI Grant Number JP25KJ1912.

\appendix
\section{Matrix elements}

In this appendix, we list up explicit matrix elements we used in the numerical calculations. 
\begin{eqnarray}
    \alpha^{(\rho)}_{n}\eqns\bar{\alpha}^{(\rho)}_n\\[6pt]
    \beta^{(\rho)}_{n}\eqns \bar{\beta}^{(\rho)}_n- \frac{\rho_{\rm I}}{2M\omega}\left[i\omega\alpha\phi_{\rm c} +\frac{\alpha\phi_{\rm c}}{2M}(\rho_{\rm H}+n)\right]\\[6pt]
    \gamma^{(\rho)}_{n} \eqns \bar{\gamma}^{(\rho)}_n-\frac{\rho_{\rm I}}{2M\omega}\left[-2i\omega\alpha\phi_{\rm c}-\frac{\alpha\phi_{\rm c}}{2M}(6\rho_{\rm H}+2\rho_{\rm I}+6(n-1)+2)\right]\\[6pt]
	\delta^{(\rho)}_{n} \eqns \bar{\delta}^{(\rho)}_n-\frac{\rho_{\rm I}}{2M\omega}\left[i\omega\alpha\phi_{\rm c}+\frac{\alpha\phi_{\rm c}}{2M}(15\rho_{\rm H}+9\rho_{\rm I}+15(n-2)+10)\right]\\[6pt]
	\epsilon^{(\rho)}_{n} \eqns \bar{\epsilon}^{(\rho)}_n-\frac{\rho_{\rm I}}{2M\omega}\left[-\frac{\alpha\phi_{\rm c}}{2M}(20\rho_{\rm H}+16\rho_{\rm I}+20(n-3)+20)\right]\\[6pt]
	\zeta^{(\rho)}_{n} \eqns -\frac{\rho_{\rm I}}{2M\omega}\left[\frac{\alpha\phi_{\rm c}}{2M}(15\rho_{\rm H}+14\rho_{\rm I}+15(n-4)+20)\right]\\[6pt]
	\eta^{(\rho)}_{n} \eqns -\frac{\rho_{\rm I}}{2M\omega}\left[-\frac{\alpha\phi_{\rm c}}{2M}(6\rho_{\rm H}+6\rho_{\rm I}+6(n-5)+10)\right]\\[6pt]
	\theta^{(\rho)}_{n} \eqns -\frac{\rho_{\rm I}}{2M\omega}\left[\frac{\alpha\phi_{\rm c}}{2M}(\rho_{\rm H}+\rho_{\rm I}+(n-6)+2)\right]
    \end{eqnarray}   
    \begin{eqnarray}
    \alpha^{(\sigma)}_{n}\eqns\bar{\alpha}^{(\sigma)}_n\\[6pt]
    \beta^{(\sigma)}_{n}\eqns \bar{\beta}^{(\sigma)}_n + \frac{\sigma_{\rm I}}{2M\omega}\left[i\omega\alpha\phi_{\rm c} +\frac{\alpha\phi_{\rm c}}{2M}(\sigma_{\rm H}+n)\right]\\[6pt]
    \gamma^{(\sigma)}_{n} \eqns \bar{\gamma}^{(\sigma)}_n + \frac{\sigma_{\rm I}}{2M\omega}\left[-2i\omega\alpha\phi_{\rm c}-\frac{\alpha\phi_{\rm c}}{2M}(6\sigma_{\rm H}+2\sigma_{\rm I}+6(n-1)+2)\right]\\[6pt]
	\delta^{(\sigma)}_{n} \eqns \bar{\delta}^{(\sigma)}_n + \frac{\sigma_{\rm I}}{2M\omega}\left[i\omega\alpha\phi_{\rm c}+\frac{\alpha\phi_{\rm c}}{2M}(15\sigma_{\rm H}+9\sigma_{\rm I}+15(n-2)+10)\right]\\[6pt]
	\epsilon^{(\sigma)}_{n} \eqns \bar{\epsilon}^{(\sigma)}_n + \frac{\sigma_{\rm I}}{2M\omega}\left[-\frac{\alpha\phi_{\rm c}}{2M}(20\sigma_{\rm H}+16\sigma_{\rm I}+20(n-3)+20)\right]\\[6pt]
	\zeta^{(\sigma)}_{n} \eqns \frac{\sigma_{\rm I}}{2M\omega}\left[\frac{\alpha\phi_{\rm c}}{2M}(15\sigma_{\rm H}+14\sigma_{\rm I}+15(n-4)+20)\right]\\[6pt]
	\eta^{(\sigma)}_{n} \eqns \frac{\sigma_{\rm I}}{2M\omega}\left[-\frac{\alpha\phi_{\rm c}}{2M}(6\sigma_{\rm H}+6\sigma_{\rm I}+6(n-5)+10)\right]\\[6pt]
	\theta^{(\sigma)}_{n} \eqns \frac{\sigma_{\rm I}}{2M\omega}\left[\frac{\alpha\phi_{\rm c}}{2M}(\sigma_{\rm H}+\sigma_{\rm I}+(n-6)+2)\right]
    \end{eqnarray}
    
    \begin{eqnarray}
    A^{(\rho)}_{n}\eqns - \frac{\rho_{\rm I}}{2M\omega}\bar{\alpha}^{(\rho)}_n\\[6pt]
    B^{(\rho)}_{n}\eqns - \frac{\rho_{\rm I}}{2M\omega}\bar{\beta}^{(\rho)}_n - \left[i\omega\alpha\phi_{\rm c} +\frac{\alpha\phi_{\rm c}}{2M}(\rho_{\rm H}+n)\right]\\[6pt]
    C^{(\rho)}_{n} \eqns - \frac{\rho_{\rm I}}{2M\omega}\left[\bar{\gamma}^{(\rho)}_n+6(\alpha \phi_{\rm c})^2\right]-\left[-2i\omega\alpha\phi_{\rm c}-\frac{\alpha\phi_{\rm c}}{2M}(6\rho_{\rm H}+2\rho_{\rm I}+6(n-1))\right]\\[6pt]
	D^{(\rho)}_{n} \eqns - \frac{\rho_{\rm I}}{2M\omega}\left[\bar{\delta}^{(\rho)}_n-15(\alpha \phi_{\rm c})^2\right]-\left[i\omega\alpha\phi_{\rm c}+\frac{\alpha\phi_{\rm c}}{2M}(15\rho_{\rm H}+9\rho_{\rm I}+15(n-2))\right]\\[6pt]
	E^{(\rho)}_{n} \eqns - \frac{\rho_{\rm I}}{2M\omega}\left[\bar{\epsilon}^{(\rho)}_n+20(\alpha \phi_{\rm c})^2\right]-\left[-\frac{\alpha\phi_{\rm c}}{2M}(20\rho_{\rm H}+16\rho_{\rm I}+20(n-3))\right]\\[6pt]
	F^{(\rho)}_{n} \eqns - \frac{\rho_{\rm I}}{2M\omega}\left[-15(\alpha \phi_{\rm c})^2\right]-\left[\frac{\alpha\phi_{\rm c}}{2M}(15\rho_{\rm H}+14\rho_{\rm I}+15(n-4))\right]\\[6pt]
	G^{(\rho)}_{n} \eqns - \frac{\rho_{\rm I}}{2M\omega}\left[6(\alpha \phi_{\rm c})^2\right]-\left[-\frac{\alpha\phi_{\rm c}}{2M}(6\rho_{\rm H}+6\rho_{\rm I}+6(n-5))\right]\\[6pt]
	H^{(\rho)}_{n} \eqns - \frac{\rho_{\rm I}}{2M\omega}\left[-(\alpha \phi_{\rm c})^2\right]-\left[\frac{\alpha\phi_{\rm c}}{2M}(\rho_{\rm H}+\rho_{\rm I}+(n-6))\right]
    \end{eqnarray}   
    \begin{eqnarray}
    A^{(\sigma)}_{n}\eqns \frac{\sigma_{\rm I}}{2M\omega}\bar{\alpha}^{(\sigma)}_n\\[6pt]
    B^{(\sigma)}_{n}\eqns \frac{\sigma_{\rm I}}{2M\omega}\bar{\beta}^{(\sigma)}_n - \left[i\omega\alpha\phi_{\rm c} +\frac{\alpha\phi_{\rm c}}{2M}(\sigma_{\rm H}+n)\right]\\[6pt]
    C^{(\sigma)}_{n} \eqns \frac{\sigma_{\rm I}}{2M\omega}\left[\bar{\gamma}^{(\sigma)}_n+6(\alpha \phi_{\rm c})^2\right]-\left[-2i\omega\alpha\phi_{\rm c}-\frac{\alpha\phi_{\rm c}}{2M}(6\sigma_{\rm H}+2\sigma_{\rm I}+6(n-1))\right]\\[6pt]
	D^{(\sigma)}_{n} \eqns \frac{\sigma_{\rm I}}{2M\omega}\left[\bar{\delta}^{(\sigma)}_n-15(\alpha \phi_{\rm c})^2\right]-\left[i\omega\alpha\phi_{\rm c}+\frac{\alpha\phi_{\rm c}}{2M}(15\sigma_{\rm H}+9\sigma_{\rm I}+15(n-2))\right]\\[6pt]
	E^{(\sigma)}_{n} \eqns \frac{\sigma_{\rm I}}{2M\omega}\left[\bar{\epsilon}^{(\sigma)}_n+20(\alpha \phi_{\rm c})^2\right]-\left[-\frac{\alpha\phi_{\rm c}}{2M}(20\sigma_{\rm H}+16\sigma_{\rm I}+20(n-3))\right]\\[6pt]
	F^{(\sigma)}_{n} \eqns \frac{\sigma_{\rm I}}{2M\omega}\left[-15(\alpha \phi_{\rm c})^2\right]-\left[\frac{\alpha\phi_{\rm c}}{2M}(15\sigma_{\rm H}+14\sigma_{\rm I}+15(n-4))\right]\\[6pt]
	G^{(\sigma)}_{n} \eqns \frac{\sigma_{\rm I}}{2M\omega}\left[6(\alpha \phi_{\rm c})^2\right]-\left[-\frac{\alpha\phi_{\rm c}}{2M}(6\sigma_{\rm H}+6\sigma_{\rm I}+6(n-5))\right]\\[6pt]
	H^{(\sigma)}_{n} \eqns \frac{\sigma_{\rm I}}{2M\omega}\left[-(\alpha \phi_{\rm c})^2\right]-\left[\frac{\alpha\phi_{\rm c}}{2M}(\sigma_{\rm H}+\sigma_{\rm I}+(n-6))\right]
    \end{eqnarray}
    \begin{eqnarray}
    \bar{\alpha}^{(\rho)}_n \eqns (\rho_{\rm H} + \rho_{\rm I})(2 + \rho_{\rm H} + \rho_{\rm I}) - (n + 1) + (n + 1)^2 + (1 + 2\rho_{\rm H}+2\rho_{\rm I})(n + 1)  \no\\
    && - 2(\rho_{\rm H} + \rho_{\rm I}) - 2\rho_{\rm I}(\rho_{\rm H} + \rho_{\rm I})-2\rho_{\rm I}(n+1) + (-\kappa^2 + \rho_{\rm I}^2) \\[6pt]
    \bar{\beta}^{(\rho)}_n \eqns -4\left[(\rho_{\rm H} + \rho_{\rm I})(2 + \rho_{\rm H} + \rho_{\rm I}) - n + n^2 + (1 + 2\rho_{\rm H}+2\rho_{\rm I})n\right]  \no\\
    && + 6(\rho_{\rm H} + \rho_{\rm I}) + 4\rho_{\rm I}(\rho_{\rm H} + \rho_{\rm I}) - l(l + 1) + 4\rho_{\rm I} n - 2n \\[6pt]
    \bar{\gamma}^{(\rho)}_n \eqns 6\left[(\rho_{\rm H} + \rho_{\rm I})(2 + \rho_{\rm H} + \rho_{\rm I}) - (n - 1) + (n - 1)^2 + (1 + 2\rho_{\rm H}+2\rho_{\rm I})(n - 1)\right]  \no\\
    && - 6(\rho_{\rm H} + \rho_{\rm I}) - 2\rho_{\rm I}(\rho_{\rm H} + \rho_{\rm I}) - 2\rho_{\rm I}(n - 1) + 6(n - 1) + 2l(l + 1) \\[6pt]
    \bar{\delta}^{(\rho)}_n \eqns -4\left[(\rho_{\rm H} + \rho_{\rm I})(2 + \rho_{\rm H} + \rho_{\rm I}) - (n - 2) + (n - 2)^2 + (1 + 2\rho_{\rm H}+2\rho_{\rm I})(n - 2)\right] \no\\
    && + 2(\rho_{\rm H} + \rho_{\rm I}) - 6(n - 2) - l(l + 1) \\[6pt]
    \bar{\epsilon}^{(\rho)}_n \eqns (\rho_{\rm H} + \rho_{\rm I})(2 + \rho_{\rm H} + \rho_{\rm I}) - (n - 3) + (n - 3)^2 + (1 + 2\rho_{\rm H}+2\rho_{\rm I})(n - 3)+ 2(n - 3)\no\\
    \end{eqnarray}
    \begin{eqnarray}
    \bar{\alpha}^{(\sigma)}_n \eqns (\sigma_{\rm H} + \sigma_{\rm I})(2 + \sigma_{\rm H} + \sigma_{\rm I}) - (n + 1) + (n + 1)^2 + (1 + 2\sigma_{\rm H}+2\sigma_{\rm I})(n + 1)  \no\\
    && - 2(\sigma_{\rm H} + \sigma_{\rm I}) - 2\sigma_{\rm I}(\sigma_{\rm H} + \sigma_{\rm I})-2\sigma_{\rm I}(n+1) + (-\kappa^2 + \sigma_{\rm I}^2) \\[6pt]
    \bar{\beta}^{(\sigma)}_n \eqns -4\left[(\sigma_{\rm H} + \sigma_{\rm I})(2 + \sigma_{\rm H} + \sigma_{\rm I}) - n + n^2 + (1 + 2\sigma_{\rm H}+2\sigma_{\rm I})n\right]  \no\\
    && + 6(\sigma_{\rm H} + \sigma_{\rm I}) + 4\sigma_{\rm I}(\sigma_{\rm H} + \sigma_{\rm I}) - l(l + 1) + 4\sigma_{\rm I} n - 2n \\[6pt]
    \bar{\gamma}^{(\sigma)}_n \eqns 6\left[(\sigma_{\rm H} + \sigma_{\rm I})(2 + \sigma_{\rm H} + \sigma_{\rm I}) - (n - 1) + (n - 1)^2 + (1 + 2\sigma_{\rm H}+2\sigma_{\rm I})(n - 1)\right]  \no\\
    && - 6(\sigma_{\rm H} + \sigma_{\rm I}) - 2\sigma_{\rm I}(\sigma_{\rm H} + \sigma_{\rm I}) - 2\sigma_{\rm I}(n - 1) + 6(n - 1) + 2l(l + 1) \\[6pt]
    \bar{\delta}^{(\sigma)}_n \eqns -4\left[(\sigma_{\rm H} + \sigma_{\rm I})(2 + \sigma_{\rm H} + \sigma_{\rm I}) - (n - 2) + (n - 2)^2 + (1 + 2\sigma_{\rm H}+2\sigma_{\rm I})(n - 2)\right] \no\\
    && + 2(\sigma_{\rm H} + \sigma_{\rm I}) - 6(n - 2) - l(l + 1) \\[6pt]
    \bar{\epsilon}^{(\sigma)}_n \eqns (\sigma_{\rm H} + \sigma_{\rm I})(2 + \sigma_{\rm H} + \sigma_{\rm I}) - (n - 3) + (n - 3)^2 + (1 + 2\sigma_{\rm H}+2\sigma_{\rm I})(n - 3)+ 2(n - 3)\no\\
    \end{eqnarray}


\bibliography{BH} 
\bibliographystyle{unsrt}
\end{document}